\documentclass{PoS}

\usepackage{wrapfig}

\usepackage{amsmath}
\usepackage{amsfonts,amsbsy}
\usepackage{amssymb}
\usepackage{appendix}
\usepackage{array}
\usepackage{multirow}
\def\be{\begin{equation}}
\def\ee{\end{equation}}
\def\bea{\begin{eqnarray}}
\def\eea{\end{eqnarray}}

\def\eq#1{{Eq.~(\ref{#1})}}
\def\fig#1{{Fig.~\ref{#1}}}

\title{IP-Sat: Impact-Parameter dependent Saturation model; revised }

\ShortTitle{IP-Sat: Impact-Parameter dependent Saturation model; revised}

\author{\speaker{Amir H. Rezaeian}%
 \\
Departamento de F\'\i sica y Centro Cient\'\i fico Tecnol\'ogico de Valpara\'\i so,  Universidad T\'ecnica
Federico Santa Mar\'\i a, 
Casilla 110-V, Valparaiso, Chile\\
E-mail: \email{Amir.Rezaeian@usm.cl}}

\author{Marat~Siddikov\\
        Departamento de F\'\i sica, Universidad T\'ecnica
    Federico Santa Mar\'\i a,
     Casilla 110-V, Valparaiso, Chile}

\author{Merijn Van de Klundert\\Physics Department, Universiteit Antwerpen, Groenenborgerlaan 171, 2020 Antwerpen, Belgium  }

\author{Raju Venugopalan\\
          Physics Department, Bldg. 510A, Brookhaven National Laboratory, Upton, NY 11973, USA
         }

\abstract{In this talk, we present a global analysis of available small-x data on inclusive DIS and exclusive diffractive processes, including the latest data from the combined HERA analysis on reduced cross sections within the Impact-Parameter dependent Saturation (IP-Sat) Model. The impact-parameter dependence of dipole amplitude is crucial  in order to have a unified description of both inclusive and exclusive diffractive processes. With the parameters of model fixed via a fit to the high-precision reduced cross-section, we compare model predictions to data for the structure functions, the longitudinal structure function, the charm structure function, exclusive vector mesons production and Deeply Virtual Compton Scattering (DVCS). Excellent agreement is obtained for the processes considered at small $x$ in a wide range of $Q^2$.
}

\FullConference{XXI International Workshop on Deep-Inelastic Scattering and Related Subjects\\
                 22-26 April, 2013\\
                 Marseilles, France}

\begin{document}

\section{Introduction}
Exclusive diffractive processes at HERA such as exclusive vector meson production or deeply virtual Compton scattering (DVCS) alongside with inclusive DIS are excellent probes of the high-energy limit of QCD. An effective field theory describing the high-energy limit of QCD is the Color Glass Condensate (CGC) \cite{mv,cgc-review1}. 
A key ingredient in particle production at small-x in the CGC approach is the universal dipole amplitude, the imaginary part of the quark-antiquark scattering amplitude on a proton or nuclear target.
A simple dipole model that incorporates the physics of saturation and models the impact parameter dependence of gluon distributions is the IP-Sat dipole model \cite{Kowalski:2003hm,watt2007,ipsat-n}.  This model for the dipole amplitude, whose form can be derived at the classical level in the CGC~\cite{mv}, contains an eikonalized gluon distribution which satisfies DGLAP evolution while explicitly maintaining unitarity. It also matches smoothly to the high $Q^2$ perturbative QCD limit. The impact parameter dependence of the amplitude  allows one to confront a large body of HERA data on exclusive diffractive processes which cannot otherwise be described simply in saturation models.  The IP-Sat model\footnote{Here, we only focus on the IP-Sat model and we do not consider the b-CGC model \cite{watt-bcgc,bcgc-n} which is an alternative impact-parameter dependent saturation model  that has been applied to many reactions including diffractive processes \cite{watt2007,watt-bcgc,pp-LR}.} attempts to approach the saturation boundary via DGLAP evolution; the eikonalization of the gluon distribution represents higher twist contributions that are becoming important at small $x$.

The main purpose of this study is to reexamine the IP-Sat model in view of  recent precise data from HERA \cite{Aaron:2009aa,Abramowicz:1900rp} and to obtain its free parameters from a fit.  Below, we summarize a few key results,  the details can be found in Ref.\,\cite{ipsat-n}. A numerical code (C++ and Fortran) for the IP-Sat dipole amplitude (with self-contained DGLAP evolution) is available for download at: 
sites.google.com/site/drarezaeian/IP-Sat.tar.gz?attredirects=$0\&$d=1.

\vspace{-0.2cm}
\section{Inclusive DIS and exclusive diffractive processes; a unified description}
\vspace{-0.2cm}
In the dipole picture, the scattering amplitude for the exclusive diffractive process $\gamma^*+p\to E+p$ with a final-state vector meson $E=J/\Psi, \phi,\rho$  or a real photon $E=\gamma$  in DVCS, can be written in terms of a convolution of the $q\bar{q}$ dipole-proton scattering amplitude $\mathcal{N}$ and the overlap wave-functions of photon and the exclusive final-state particle $\Psi_{E}^{*}\Psi$  \cite{watt2007,ipsat-n}, 
\begin{equation} \label{am-i}
  \mathcal{A}^{\gamma^* p\rightarrow Ep}_{T,L} (x,Q, \Delta)= \mathrm{2i}\,\int d^2\vec{r}\int_0^1 dz \int d^2\vec{b}\;(\Psi_{E}^{*}\Psi)_{T,L}\;\mathrm{e}^{-\mathrm{i}[\vec{b}-(1-z)\vec{r}]\cdot\vec{\Delta}}\mathcal{N}\left(x,r,b\right), 
\end{equation}
where $\vec{\Delta}^2=-t$ with $t$ being the squared momentum transfer, $r$ and b denote the dipole transverse size and impact-parameter of the collision, respectively. The differential cross-section for the exclusive diffractive process can be then given,
\begin{equation}
 \frac{d\sigma^{\gamma^* p\rightarrow Ep}_{T,L}}{d t}  = \frac{1}{16\pi}\left\lvert\mathcal{A}^{\gamma^* p\rightarrow Ep}_{T,L}\right\rvert^2\;(1+\beta^2),
  \label{vm}
\end{equation}
where the factor $(1+\beta^2)$ takes into account the real part of amplitude in \eq{am-i} and $\beta$ is the ratio of the real to imaginary parts of the scattering amplitude,
 \begin{equation} \label{eq:beta}
  \beta = \tan\left(\frac{\pi\lambda}{2}\right), \quad\text{with}\quad \lambda \equiv \frac{\partial\ln\left(\mathcal{A}_{T,L}^{\gamma^* p\rightarrow Ep}\right)}{\partial\ln(1/x)}.
\end{equation}
The total deeply inelastic cross-section for a given $x$ and $Q^2$ can be obtained from \eq{am-i},
\begin{equation}\label{gp}
  \sigma_{L,T}^{\gamma^*p}(Q^2,x) =Im   \mathcal{A}^{\gamma^* p\rightarrow Ep}_{T,L} (x,Q, \Delta=0).
\end{equation}
The proton structure function $F_2$, the longitudinal structure function $F_L$  and reduced cross-section $\sigma_{r}$  can be then written in terms of the total $\gamma^{\star}p$ cross-section as
\begin{eqnarray}
F_2(Q^2,x) &=& \frac{Q^2}{4\pi^2\alpha_{EM}} 
\left[\sigma_L^{\gamma^*p}(Q^2,x)+\sigma_T^{\gamma^*p}(Q^2,x)\right],\label{f2}\\
F_L(Q^2,x) &=& \frac{Q^2}{4\pi^2\alpha_{EM}}\sigma_L^{\gamma^*p}(Q^2,x).  \label{FL}\\
\sigma_{r}\left(x,y,Q^{2}\right)&=&F_{2}\left(x,Q^{2}\right)-\frac{y^{2}}{1+(1-y)^{2}}F_{L}\left(x,Q^{2}\right),\
\end{eqnarray}
where $y=Q^2/(sx)$ is the inelasticity variable and $\sqrt{s}$ denotes the center of mass energy in $ep$ collisions. In the above expression, we neglected the contribution of the $Z$ boson which is important only at very large $Q^2$.

\begin{figure}[t]       
\includegraphics[width=0.9\textwidth,clip]{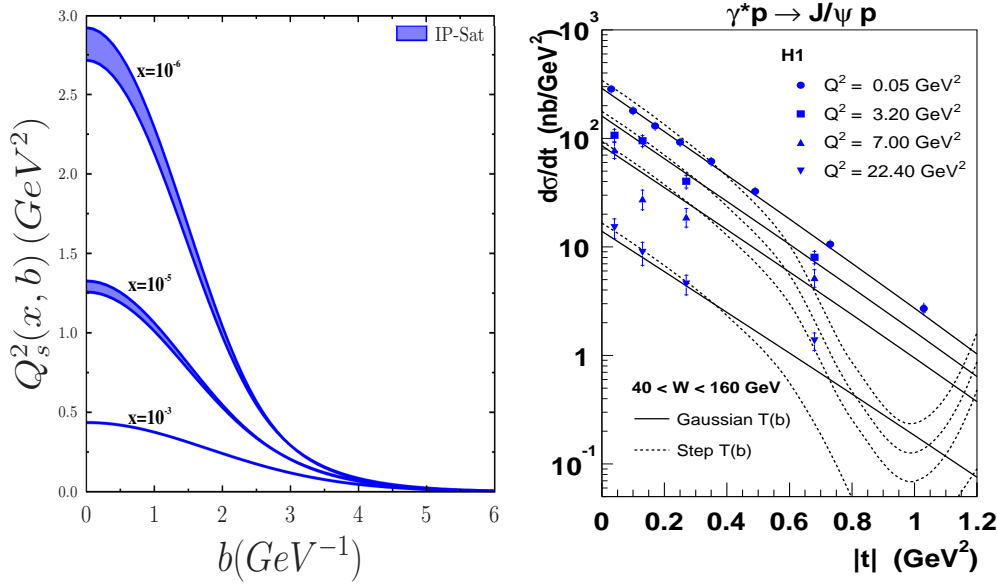}   
\caption{Left: The saturation scale as a function of the impact-parameter $b$ for various fixed values of $x$. Right: Differential $J/\Psi$ cross-section as a function of $|t|$ obtained by a a Gaussian or a step function for the impact-parameter profile of the dipole amplitude. The plots are from Refs.~\cite{watt2007,ipsat-n}.}
\label{f1}        
\end{figure}

The common ingredient of the cross-sections in DIS, exclusive diffractive vector meson production and  DVCS is the universal $q\bar{q}$ dipole-target amplitude.  As seen in Eqs.\,(\ref{am-i}, \ref{vm}), the impact-parameter dependence of the dipole amplitude is crucial for describing  exclusive diffractive processes. For the total cross-section, the effect of the impact-parameter dependence of the dipole amplitude is not especially important and the $b$-dependence can be effectively incorporated by treating it as
 a step function and adjusting the overall normalization. In this way, one can still find a good fit for the structure functions and total DIS cross-section. However, a consequence of a trivial $b$-dependence leads to a pronounced dip in the $t$-distribution of vector meson production at low $|t|$. This is not observed in data and can therefore be ruled out, see \fig{f1}. A simple $b$-dependence for the dipole amplitude is obtained by combining the Glauber-Mueller form \cite{Kowalski:2003hm,watt2007,ipsat-n}  of the amplitude
\bea
\mathcal{N}\left(x,r,b\right)  &=&1-\exp\left(-\frac{\pi^{2}r^{2}}{2N_{c}}\alpha_{s}\left(\mu^{2}\right)xg\left(x,\mu^{2}\right)T_{G}(b)\right), \label{ip-sat} 
\eea
with a Gaussian impact parameter profile
\bea
T_{G}(b)&=& \frac{1}{2\pi B_G}\exp\left(-b^2/2B_G\right) \label{ip-b} \, ,\
\eea
 In the above, $xg\left(x,\mu_{0}^{2}\right)$ is the gluon density evolved up to  the scale $\mu$ with LO DGLAP gluon evolution.   
  The parameter $B_G$ will be fixed with experimental data for exclusive $J/\Psi$ production. We take the corresponding one loop running-coupling value of $\alpha_s$ with $\Lambda_{\text{QCD}}=0.156$ GeV fixed by the experimentally measured value of $\alpha_s$ at the $Z^0$ mass. The contribution from bottom quarks is neglected. As in the original IP-Sat model, the scale $\mu^2$ is related to the dipole transverse size by
\begin{equation}
 \mu^{2}=4/r^{2}+\mu_{0}^{2}, \label{c}
\end{equation}
and the initial gluon distribution at the scale $\mu_0^2$ is taken to be  
\begin{equation}
 xg\left(x,\mu_{0}^{2}\right) =A_{g}\,x^{-\lambda_{g}}(1-x)^{5.6} \label{g}.
\end{equation}
The parameters $A_{g},\lambda_{g}, \mu_{0}^{2}$ and $B_G$ are the only free parameters of our model which will be fixed by a fit to the reduced cross-section. In \fig{f1}(left), we show the impact-parameter dependence of the saturation scale in the IP-Sat model. We should stress that in the dipole approach, the impact-parameter profile of the saturation scale is closely related to the $t$-distribution of the exclusive diffractive processes, as demonstrated in  \fig{f1} right panel.  

\begin{figure}
  \includegraphics[width=0.45\textwidth,clip]{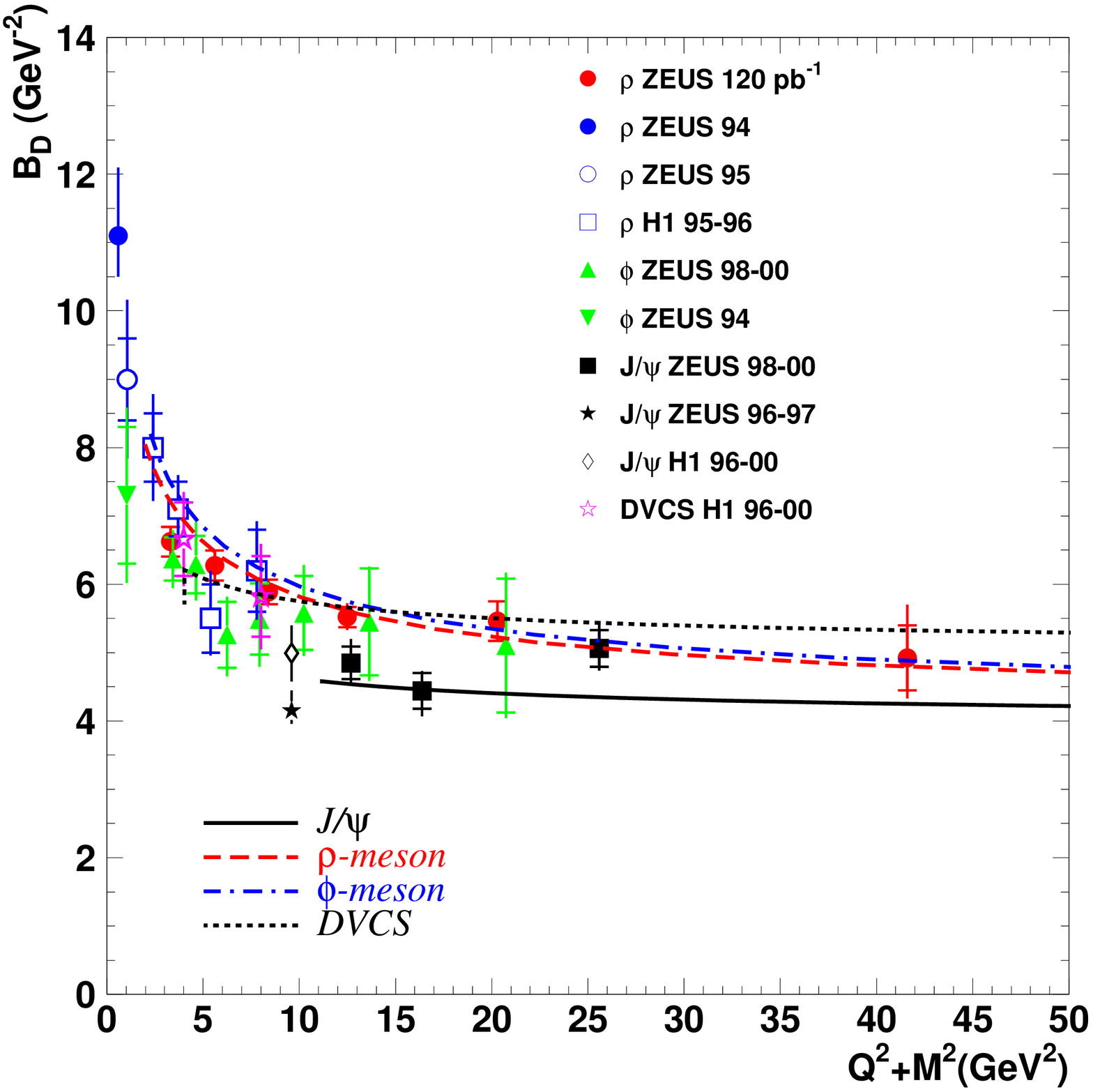}
\includegraphics[width=0.45\textwidth,clip]{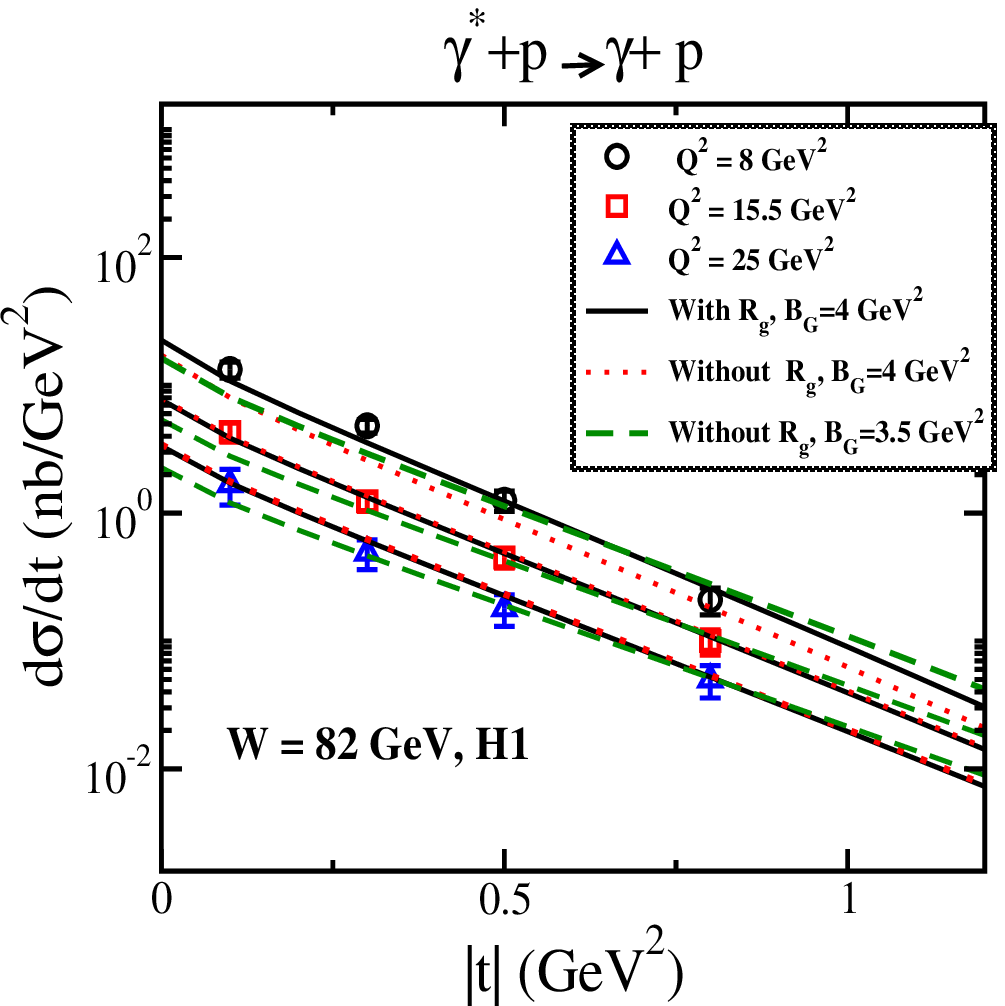}
  \caption{Left: A compilation of the value of the slope $B_D$ of $t$-distribution of exclusive vector-meson electroproduction and DVCS processes, as a function of $Q^2+M^2_V$. Right: Differential DVCS cross-section as a function of $|t|$ with or without the inclusion of the skewedness factor $R_g$ defined in. The experimental data are from \cite{Chekanov:2007zr,Aaron:2009ac}. The left plot is taken from Ref.~\cite{ipsat-n}. }
  \label{f22}
\end{figure}

For exclusive diffractive processes, one should also incorporate the skewedness effect due to the fact that the gluons attached to the $q\bar{q}$ can carry different light-cone fractions $x,x^{\prime}$ of proton. At NLO level, in the limit that $x^{\prime}<<x<<1$ and small $|t|$,  the skewedness effect \cite{ske} can be effectively accounted for by simply multiplying the gluon distribution $xg(x,\mu^2)$ by a factor $R_g$ defined via \cite{ske}, 
\begin{equation} \label{eq:Rg}
  R_g(\gamma) = \frac{2^{2\gamma+3}}{\sqrt{\pi}}\frac{\Gamma(\gamma+5/2)}{\Gamma(\gamma+4)}, \quad\text{with}\quad \gamma \equiv \frac{\partial\ln\left[xg(x,\mu^2)\right]}{\partial\ln(1/x)}.
\end{equation}
In obtaining the factor $R_g$, it was assumed that the diagonal gluon density of proton has a power-law form of $xg(x)\sim x^{-\gamma}$ which makes sense at small-x and is consistent with our parametrization in \eq{g}.  There is uncertainty with regard to how one incorporates the skewedness correction at small $x$, and the factor $R_g$ should be regarded as a phenomenological estimate. Nevertheless, the gluon distribution is mainly determined from the reduced cross-section (or structure functions) alone; the choice of $R_g$ will only slightly affect the parametrization of the former.  We fixed the width of proton impact-parameter profile $B_G$ in \eq{ip-b} via a fit to the slope of the $t$-distribution of the $J/\Psi$ mesons and we found $B_G=4\,\text{GeV}^{-2}$. As it is seen in \fig{f22} (left), the experimental errors for the slope of $t$-distribution of exclusive diffractive vector mesons and DVCS  (denoted by $B_D$) are rather large.  We estimated that the uncertainties of the value of $B_G$ is about $0.4\div 0.5\,\text{GeV}^{-2}$. Although, the inclusion of the skewedness effect improves our description of diffractive exclusive processes(indicating the importance of higher order corrections), but some part of effect may be also absorbed into our uncertainties in extracting the parameter $B_G$. The effect of the inclusion of $R_g$  is shown in \fig{f22} (right) for the case of DVCS production where the uncertainties with respect to the overlap wavefunction and charm mass dependence is significantly less compared to vector mesons production.

At large values of $M^2+Q^2$ (with $M$ being the vector meson mass),  we are in the color transparency regime and the main contribution to \eq{am-i} comes from small dipole sizes. Therefore the $t$-distribution at small dipole sizes can be approximately determined by the Fourier transform of $T_{G}(b)$ in \eq{ip-b},
\begin{equation}
 \frac{d\sigma^{\gamma^* p\rightarrow Ep}_{T,L}}{d t}\approx e^{-B_G|t|}, \label{bd}
\end{equation}
which is fully supported by the experimental data shown in \fig{f22} (left).  Note that in general, $B_D\neq B_G$ because the 
Fourier transform of the $b$-distribution in the IP-Sat exclusive vector-meson amplitude is not a simple exponential in $|t|$. 
Therefore, at a fixed virtuality, the typical dipole size is bigger for lighter vector meson and consequently the validity of the asymptotic expression in \eq{bd}  is postponed to a higher virtuality.  It is seen \fig{f22} (left)  that indeed at large $Q^2+M^2_V$  the value of $B_D$  tends to saturate to a universal value mainly determined by the interaction area and impact-parameter profile of proton. 

\begin{figure}[t]       
\includegraphics[width=0.46\textwidth,clip]{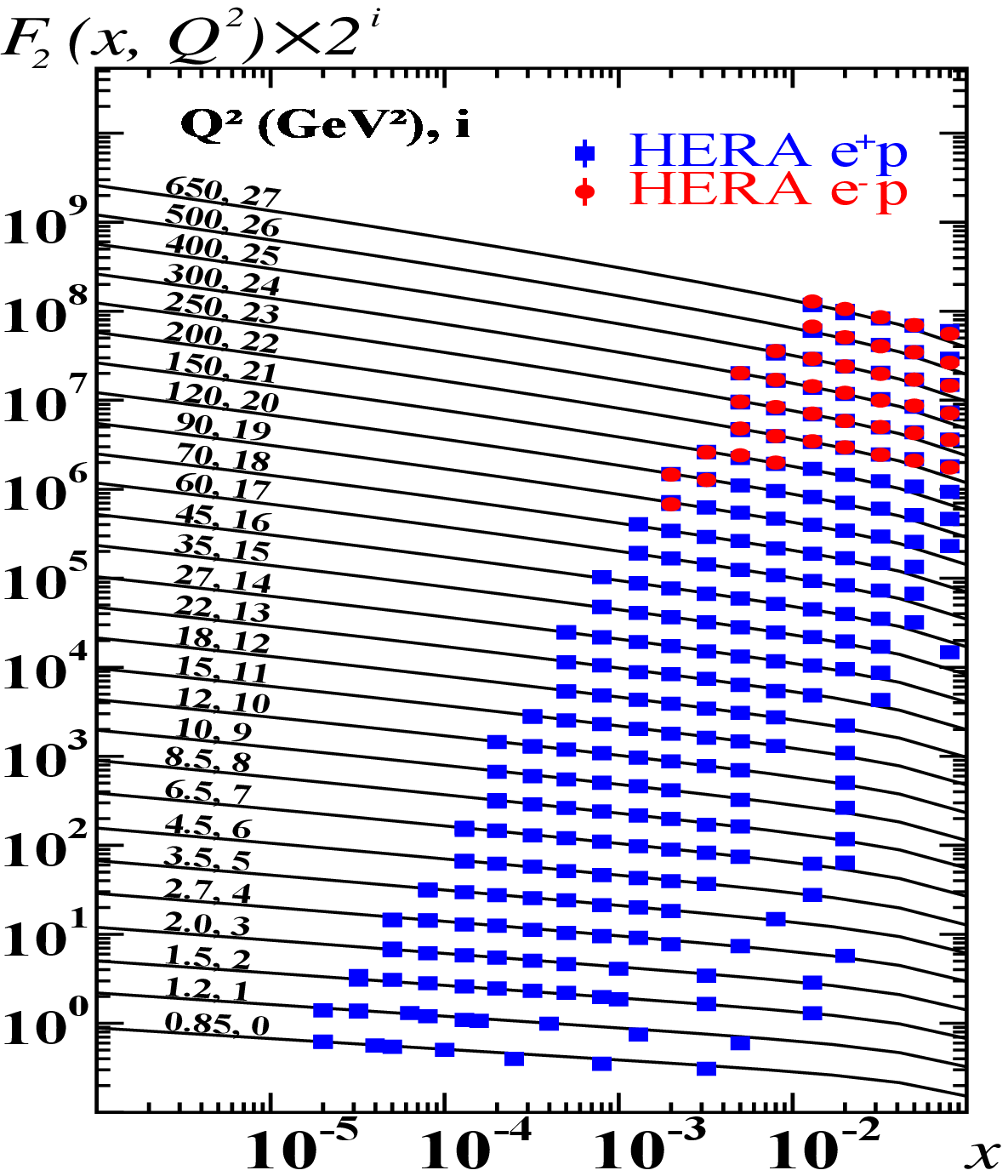}     
\includegraphics[width=0.5\textwidth,clip]{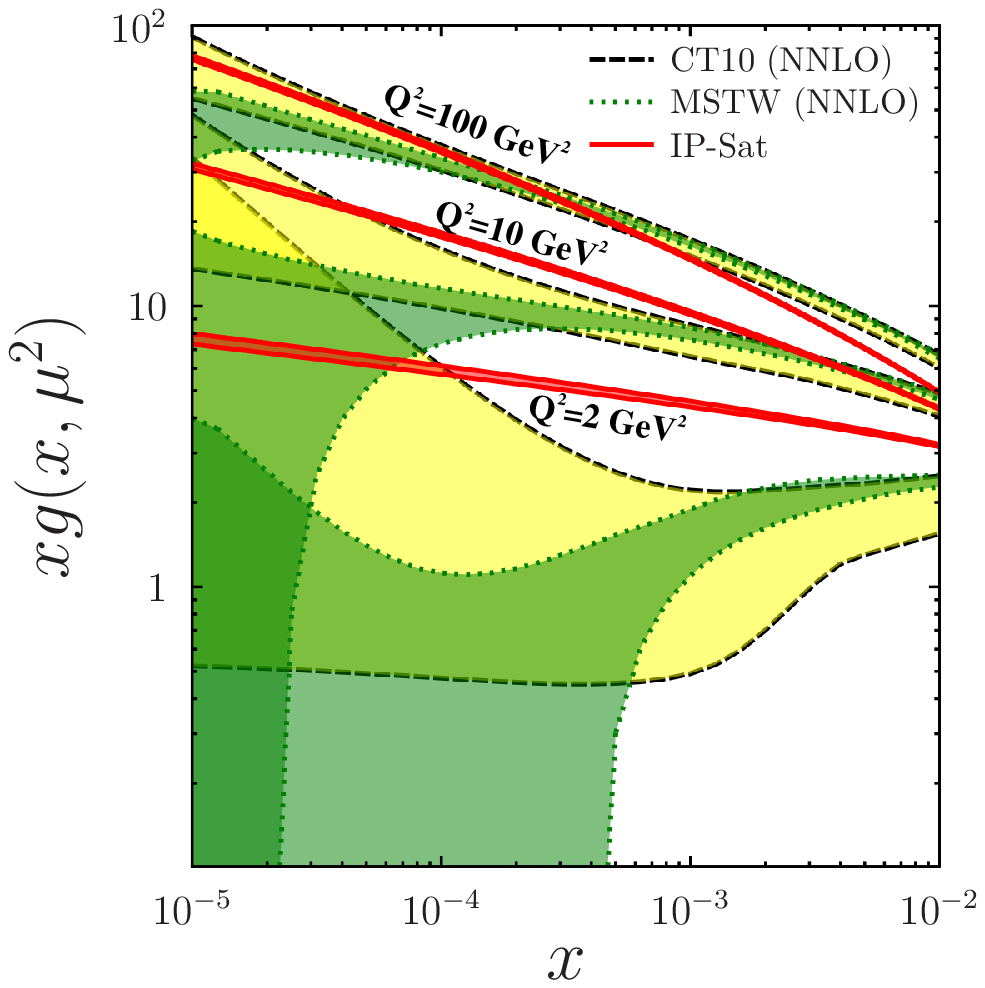}   
\caption{ Left: The gluon structure function as a function of $x$ for various fixed values of virtuality $Q^2$ extracted in the dipole saturation model (IP-Sat),  CT10 (NNLO) \cite{Lai:2010vv} and MSTW 2008 (NNLO) \cite{mstw}.  
The corresponding theoretical uncertainties are represented with bands between solid, dashed and dotted lines for IP-Sat, CT10 and MSTW, respectively. Right: The gluon structure function $xg\left(x, \mu^2(r)\right)$ as a function of dipole transverse size $r$ for various fixed values of $x$.  The  plots are from Ref.~\cite{ipsat-n}.}
\label{f33}           
\end{figure}     

With the parameters of the IP-Sat model extracted from the $\chi$-squared fit to the reduced inclusive DIS cross-section, we then compute the structure functions $F_2(x,Q^2)$ using Eqs.\,(\ref{f2}, \ref{ip-sat}) and compare to the combined HERA data sets in \fig{f33} (left). For comparison of our results with other observables at HERA, see Ref.\,\cite{ipsat-n}. It is remarkable that with only 4 parameters fixed to reduced cross-section, our model gives excellent description of almost all available data on inclusive and exclusive diffractive processes at HERA at small-x ($x\le 10^{-2}$).  

In \fig{f33} (right), we compare the gluon structure function  at various fixed values of virtuality $Q^2$ obtained from the IP-Sat model and the leading twist collinear factorization approach with NNLO DGLAP evolution,  namely CT10 \cite{Lai:2010vv} and MSTW 2008 \cite{mstw}. The bands for CT10 and MSTW correspond to uncertainties in obtaining a fit from global data analysis, while in the IP-Sat model the uncertainties are mainly due to our freedom to choose different values for the charm quark mass in the range $m_c=1.27 \div 1.4$ GeV. At low virtualities and low $x$, we are in the saturation regime and we observe our gluon distributions to be significantly different from those obtained from the leading twist perturbative computations and significantly more stable. We recall that the number of free parameters in our model is significantly less than the standard  collinear factorization approach, moreover, we have taken only small-x data at HERA into our analysis where we expect our formalism to be reliable. Consequently, the parameters of our model are better constrained compared to the standard pQCD approach, leading to more stable results with smaller errors. 
At large virtualities, the saturation effects become irrelevant and our approach approximately matches the standard perturbative formalism. The small differences seen at high virtualities are mainly due to the fact that we used LO DGLAP evolution without including quark degrees of freedom, while quark evolution contributions were included in the perturbative leading twist results shown in \fig{f33}.

 Other key features of our novel fit are the preferred lower values for the light quark masses $m_u\approx 0$ and also positive value for the parameter $\lambda_g>0$ in \eq{g} which are in sharp contrast with the old fit in Refs.\,\cite{Kowalski:2003hm,watt2007}.  The IP-Sat model has been intensively applied to various reactions including heavy ion collisions \cite{Schenke:2012wb}. However, the parameters employed in these studies were determined from data from H1 and ZEUS predating the combined data sets for the proton. It remains to be seen what the impact of the new fits are on final state observables in heavy ion collisions.

\end{document}